\documentclass[12pt]{amsart}
\usepackage{amsfonts,amssymb,amsmath,amscd}
\usepackage{epsf}

\def\rx#1{~(\ref{#1})}
\def\ex#1{Eq.~(\ref{#1})}
\def\rf#1{Ref.~\cite{#1}}

\def\prion{ \prod_{i=1}^n }

\def\bx{ \mathbf{x} }
\def\by{ \mathbf{y} }
\def\bfz{ \mathbf{z} }
\def\byxv#1{ (\by,\bx)_{#1} }
\def\byxv#1{ \by_{[#1]}\bx }
\def\byxi{ \byxv{i} }
\def\byximo{ \byxv{i-1} }

\def\xP{ P }

\def\Dlvv#1#2{ \Delta_{#1}(#2) }
\def\Dliv#1{ \Dlvv{i}{#1} }
\def\DliP{ \Dliv{\xP} }

\def\DlijW{ \Dliv{\partial_j W} }

\def\strD{ \str (\partial D)^{\wedge 2n} }

\def\sg{ \sigma }
\def\sgin{ \sg^{-1} }
\def\sgv#1{ \sg(#1) }

\def\sgti{ \sgv{2i} }
\def\sgtio{ \sgv{2i+1} }
\def\Sv#1{ S_{#1} }
\def\sSv#1{ \sg\in\Sv{#1} }
\def\ssSv#1{ \sum_{\sSv{#1}} }
\def\ssSvmos#1{ \ssSv{#1}\mos }

\def\Sn{ \Sv{n} }
\def\ssSnmos{ \ssSvmos{n} }

\def\ssStnmos{ \ssSvmos{2n} }

\def\mos{ (-1)^{\sgn(\sg)} }

\def\sgn{ {\mathop{{\rm sign}}\nolimits} }

\def\prtzvv#1#2{ \frac{\partial{#1}}{\partial z_{#2} } }
\def\prtzDiv#1{ \prtzvv{D_i}{#1} }

\def\prtzDistj{ \prtzDiv{j} }
\def\prtzDistk{ \prtzDiv{k} }

\def\xC{ C }
\def\xCvv#1#2#3{ \xC_{#1}[#2,#3] }
\def\xCzvv#1#2#3{ \xCvv{#1}{z_{#2}}{z_{#3}} }
\def\xCziv#1#2{ \xCzvv{i}{#1}{#2} }
\def\xCzitio{ \xCziv{\sgti}{\sgtio} }
\def\xCzijk{ \xCziv{j}{k} }

\def\xCiv#1#2{ \xCvv{i}{#1}{#2} }

\def\IDv#1{ I_{#1} }
\def\IDW{ \IDv{W} }

\def\Frob{Frobenius}
\def\trv#1{ \tr_{#1} }
\def\trF{ \trv{\mathrm{F}} }
\def\trb{ \trv{\mathrm{b}} }
\def\trx{ \trv{\bx} }

\def\JR{ \mathrm{J} }
\def\JRv#1{ \JR_{#1} }
\def\JRW{ \JRv{W} }
\def\JRWv#1{ \JRv{W(#1)} }
\def\JRWx{ \JRWv{\bx} }

\def\xAcl{ A_{\mathrm{cl} } }
\def\xAop{ A_{\mathrm{op} } }
\def\scpr#1#2{ (#1,#2) }
\def\corr#1{ \langle #1 \rangle }

\def\xp{ p }
\def\xO{ O }

\def\Dwn{ (\partial D)^{\wedge n} }

\def\Id{ \mathrm{Id} }

\def\rhs{r.h.s.}

\def\cnv#1{ {#1}^{*} }
\def\Acn{ \cnv{A} }

\def\cmult{comultiplication}

\def\xmtr#1{ ||#1|| }
\def\xrw#1{ (#1) }

\def\part{\partial}

\newcommand{\Tr}{{\rm Tr}}
\newcommand{\str}{{\rm STr}}

\newcommand{\eps}{\epsilon}

\newcommand{\CC}{{\mathbb C}}
\newcommand{\PP}{{\mathbb P}}

\newcommand{\ZZ}{{\mathbb Z}}
\newcommand{\NN}{{\mathbb N}}

\newcommand{\Hom}{{\rm Hom}}

\newcommand{\End}{{\rm End}}
\newcommand{\Ext}{{\rm Ext}}

\newcommand{\cA}{{\mathcal A}}

\newcommand{\cD}{{\mathcal D}}

\renewcommand{\O}{{\mathcal O}}

\newcommand{\ra}{\rightarrow}

\newcommand{\op}{\oplus}
\newcommand{\ot}{\otimes}

\newcommand{\id}{{\rm id}}
\newcommand{\Ho}{{Hochschild}}
\newcommand{\HH}{{\rm HH}}
\newcommand{\cM}{{\mathcal M}}
\newcommand{\cN}{{\mathcal N}}
\newcommand{\cT}{{\mathcal T}}
\newcommand{\tr}{{\rm tr}}

\begin{document}

\title[Relation between open and closed strings]{On the relation between open and closed topological strings}
\author{Anton Kapustin}{ \hspace{3mm} }
\address{California Institute of Technology\\
Department of Physics\\
Pasadena, CA 91125}
\email{kapustin@theory.caltech.edu}
\author{Lev Rozansky}{ \hspace{3mm} }
\address{University of North Carolina\\
Department of Mathematics\\
Chapel Hill, NC 27599}
\email{rozansky@math.unc.edu}

\begin{abstract}
We discuss the relation between open and closed string correlators using topological string theories as a toy model. We propose
that one can reconstruct closed string correlators from the open ones by considering the
\Ho\ cohomology of the category of D-branes. We compute the \Ho\ cohomology
of the category of D-branes in topological Landau-Ginzburg models and partially verify the conjecture
in this case.

\end{abstract}

\maketitle

\vspace{-5in}

\parbox{\linewidth}
{\small\hfill \shortstack{CALT-68-nnnn}}

\vspace{5in}

\section{Introduction and Summary}

There exist different viewpoints on the question whether closed or open strings are more fundamental. The more popular viewpoint is that closed strings are simpler (because one
does not have to make a choice of a boundary condition) and therefore more fundamental. In this
view, the central problem is to classify D-branes and construct open-closed string correlators
for a given closed string theory. In practice, complete classification is possible only in
very simple (mostly topological) examples. The opposite viewpoint is supported by the
observation that closed String Field Theory is vastly more complicated than the open one.
While in the open-string case the classical action is cubic~\cite{Witten}, in the closed-string case
it is non-polynomial~\cite{SaaZwi,KKS}.
To write down the open SFT action, one has to specify an associative
product on the space of states, a differential (i.e. a BRST operator), and an invariant scalar product. Deformations of the closed-string background change these data, so there is a map from the space of closed-string states to the space of infinitesimal deformations of the open string theory.

One difficulty with the first viewpoint is that it is not clear if the spectrum and properties
of D-branes are uniquely determined by the closed string theory. In other words, there may exist
D-branes which are perfectly consistent by themselves, but mutually incompatible, in the
sense that it is not possible to define states which correspond to open strings stretched between
two different branes. If one regards a D-brane as a boundary condition for a string worldsheet,
such a situation may appear absurd, but one must remember that up to now there does not exist a precise definition of the notion of a ``boundary condition'' at the quantum level. Instead, one
characterizes D-branes as solutions of a complicated set of conditions (see Refs.~\cite{Lazaxi,Lazrec}), the most nontrivial of which is the so-called Cardy condition.
With such an axiomatic definition of a D-brane, the situation
described above is not ruled out.

If one adopts the second viewpoint, then the central problem is to construct closed string correlators
from the open ones. As mentioned above, closed string states are
related to infinitesimal deformations of the open-string theory. Suppose for simplicity that there is only a single D-brane in the theory, so that all the information is encoded in an associative
algebra $A$ equipped with a BRST differential $Q$ of ghost number one and an invariant scalar product. The pair
$(A,Q)$ is called a differential graded algebra (DG-algebra) by mathematicians. Equivalence classes of deformations of these data are described by a certain cohomology theory (\Ho\ cohomology of $(A,Q)$ with coefficients in itself). The simplest conjecture is that the space of physical closed-string states
{\it is isomorphic} to the \Ho\ cohomology of $(A,Q)$.

At this point we must be more specific about the kind of string theory we are
talking about. In this paper we will only discuss topological string theories.
Let us recall how they are constructed. The starting point is a unitary $N=2$ $d=2$
field theory which can be twisted to a topological field theory. To get
a string theory, one has to couple it to topological gravity. On the level
of the space of states, this operation is very simple~\cite{topgravWi,Getz2}:
each state of the TFT gives rise to an infinite sequence of states of increasing
ghost number. The first state in this sequence is called a gravitational primary,
and the rest are called gravitational descendants. Tree level correlators of primaries
can be computed in the TFT, i.e. the coupling to topological gravity plays no
role for these correlators. In this paper we will only discuss such correlators,
and therefore topological gravity will play no role. The precise conjecture we
are making is that for topological strings the spectrum of gravitational primaries is given by the \Ho\ cohomology of the category of topological D-branes.

This conjecture is very appealing, because many structures of the open-closed string theory are then automatic. For example, the
Hoch\-schild cohomology of any DG-algebra is itself a supercommutative algebra, which may be
identified with the algebra of observables in the closed-string TFT. It also has a Lie-type bracket
of degree $-1$,
in agreement with the findings of Refs.~\cite{Zwiebach,WZ,LZ,Getz1,PS2,LZ2}.
Further, consider the cohomology of $Q$, which
may be regarded as the space of physical open-string states. It turns out that this space
has a natural structure of a module over the \Ho\ cohomology of $(A,Q)$, and this allows one to define the bulk-boundary map. This will be discussed in more detail below.
Many axioms of open-closed TFT (see Ref.~\cite{Lazaxi}) are then easily verified.\footnote{The Cardy condition
is an exception in this regard. It seems to be a generalization of the
Hirzebruch-Riemann-Roch theorem, and its validity is not at all obvious.}

The assumption that there is only one D-brane is unrealistic. For example, given a D-brane $M_0$, one may consider direct sums of several copies of $M_0$, as well as more complicated bound
states. If there are many possible D-branes, then
one has to take into account open strings with different boundary conditions on the two ends.
It is convenient to think of a D-brane as an object of an additive (in fact $\CC$-linear)
category, and of the space of open strings between two D-branes as the space of morphisms. Then the algebra of open string states for a particular D-brane is its endomorphism algebra. BRST
operators give rise to differentials on all spaces of morphisms, so one is actually dealing with
a differential graded category (DG category). The grading is given by the ghost number. To get
the space of physical open-string states between any two D-branes, one has to compute
the cohomology of the BRST operator on the space of morphisms.
There is a notion of \Ho\ cohomology which classifies equivalence
classes of deformations of such categories. Essentially, one lumps together all the objects
in the category into a single ``total object'' and considers the \Ho\ cohomology
of its endomorphism algebra. This is equivalent to thinking about a $\CC$-linear additive category as
an ``algebra with many objects.''
It is tempting to conjecture that the space of physical closed strings is isomorphic to the
\Ho\ cohomology of the category of D-branes. A heuristic argument for this is
explained in the end of Section~\ref{defin}. Again, the space of physical open-string
states between any two D-branes is naturally a module over the \Ho\ cohomology, so the
bulk-boundary maps are automatic.

Even when there are many possible D-branes, one can often find a D-brane
$M_0$ such that all other D-branes can be obtained as bound states of several copies of $M_0$
and its anti-brane. In mathematical terms, this means that the category of D-branes is
equivalent to the category of modules of some kind over the endomorphism algebra (which is
actually a DG-algebra) of $M_0$. Then the \Ho\ cohomology of the category of D-branes
coincides with the \Ho\ cohomology of the endomorphism algebra of $M_0$. In the physical case, the role of $M_0$ can be played by a space-filling brane.

One example where this prescription for reconstructing closed strings is known to work is the topological B-model of a Calabi-Yau manifold $X$. The algebra of closed string states is given by
\begin{equation}\label{Xhoch}
\op_{p,q} H^p(\Lambda^q TX).
\end{equation}
The category of D-branes is believed to be equivalent to the bounded derived category of $X$
denoted $D^b(X)$. More precisely, it is a DG-category whose derived category is believed
to be equivalent to $D^b(X)$. The appropriate cohomology to compute is the \Ho\
cohomology of the sheaf of algebras $\O_X$,
and one can show that the latter is isomorphic to Eq.~(\ref{Xhoch})~\cite{Swan}.
This is discussed in more detail in the next section.

In this paper we study another class of topological string theories: topological
Landau-Ginzburg models. The closed string sector has been described by C.~Vafa~\cite{Vafa}.
A simple description of the category of D-branes in LG models has been
proposed by M.~Kontsevich and derived from physical considerations in Refs.~\cite{KLi1,KLi2,Letal}
(see also Ref.~\cite{Laz}). It turns out that the category of D-branes can be thought of
as the category of CDG-modules over a certain commutative CDG-algebra, where CDG stands
for ``curved differential graded.'' (This notion is explained in detail in Ref.~\cite{KLi2}
and will be recalled below.) For LG models on smooth spaces many correlators have been computed~\cite{KLi2}. In this paper we compute the \Ho\ cohomology
of the category of Landau-Ginzburg branes and show that in this way we can recover the
closed string space of states, together with its algebra structure and scalar product, as
described in Ref.~\cite{Vafa}, as well as some open-closed correlators,
as described in Ref.~\cite{KLi2}.

One may also consider LG models on orbifolds. Such models are particularly important because
they provide an alternative description of certain Calabi-Yau sigma-models (these are so-called
Gepner models~\cite{Gepner}). The closed-string spectrum for LG orbifolds has been described
by K.~Intriligator and C.~Vafa~\cite{IV}, but its interpretation in mathematical terms
has been lacking. We check in several examples that the \Ho\ cohomology
of the category of D-branes on LG orbifolds reproduces the results of Ref.~\cite{IV} and others.
Our examples include the Gepner models for Fermat-type hypersurfaces in projective spaces.
A nice feature of the \Ho\ cohomology approach
is that the 3-point closed-string correlators (i.e. the Yukawa couplings) and the bulk-boundary
maps come out automatically.

These results provide evidence that the conjectural identification of the closed string sector
with the \Ho\ cohomology of the category of D-branes is correct, at least for
topological strings, and that certain simple open-closed correlators can be computed using
the algebraic structure of the \Ho\ complex. Hopefully, multi-point and
higher-genus correlators can also be extracted from the \Ho\ complex. The technique based on the \Ho\ complex could be useful for
the computation of the space-time superpotential for superstring compactifications with D-branes
based on Gepner models~\cite{Gepner}. These open-closed superstring backgrounds have been studied in
many papers, see e.g. Refs.~\cite{RS,GutSa,Bru1,DD,Bru2,Diac,Horec}.

\section{Definitions of the \Ho\ cohomology}\label{defin}

We begin by recalling various definitions of the \Ho\ cohomology of algebras,
DG-algebras, and affine and projective varieties. We will pick the most convenient
definition and in the next section generalize it to the case of CDG algebras relevant
for the Landau-Ginzburg models.

Let $A$ be an associative algebra over $\CC$. The \Ho\ cochain complex (with coefficients
in $A$) is the sequence of vector spaces
$$
C^n(A)=\Hom_\CC(A^{\ot n},A),\quad n=0,1,\ldots,
$$
equipped with a differential $\delta:C^n(A)\ra C^{n+1}(A)$ defined by the equation
\begin{multline}\label{Hoass}
(\delta f)(a_1,\ldots,a_{n+1})=a_1 f(a_2,\ldots,a_n)\\
+\sum_{i=1}^n (-1)^i f(a_1,\ldots,a_{i-1},
a_i a_{i+1},a_{i+2},\ldots,a_n)\\
+(-1)^{n+1} f(a_1,\ldots,a_n) a_{n+1}.
\end{multline}
The cohomology of $\delta$ in degree $n$ will be denoted $\HH^n(A)$ and called the \Ho\ cohomology of $A$ (with coefficients in $A$). The more standard notation for it is $\HH^n(A,A)$.
Each 2-cocycle $f(a_1,a_2)$ defines an infinitesimal deformation of the associative product
on $A$. That is, if we define a new product by
$$
a\star b=ab+t f(a,b),\quad t\in \CC,
$$
it will be associative to linear order in $t$ if and only if $\delta f=0$.
Trivial infinitesimal deformations (i.e. infinitesimal deformations which lead to an isomorphic algebra) are classified by 2-coboundaries, i.e. 2-cocycles of the form $\delta g$ for some 1-cochain $g(a)$.
Thus $\HH^2(A)$ classifies nontrivial
deformations of the associative algebra structure on $A$. One can give a similar interpretation to the total \Ho\ cohomology $\HH^*(A)$: it classifies infinitesimal deformations of $A$
in the class of $A_\infty$ algebras, associative algebras being a very special case of $A_\infty$
algebras~\cite{PS}.

If $A$ is a $\ZZ$-graded or $\ZZ_2$-graded algebra, the \Ho\ complex is defined somewhat
differently.
Let $A_p$ be the degree-$p$ component of $A$, so that $A_p\cdot A_q\subset A_{p+q}$.
An element $f$ of $C^n(A)$ is said to have an internal degree $p$ if
$$
f(a_1,\ldots , a_n)\in A_{p+k_1+\cdots+k_n}
$$
whenever $a_i\in A_{k_i}$. Thus each vector space $C^n(A)$ is graded by the internal degree,
and we define the total degree of an element of $C^*(A)$ to be the sum of $n$ and the internal
degree. The \Ho\ complex is graded by the total degree, and the differential is given by
\begin{multline}\label{Hoassg}
(\delta f)(a_1,\ldots,a_{n+1})=(-1)^{a\cdot f}\ a_1 f(a_2,\ldots,a_n)\\
+\sum_{i=1}^n (-1)^i f(a_1,\ldots,a_{i-1},
a_i a_{i+1},a_{i+2},\ldots,a_n)\\
+(-1)^{n+1} f(a_1,\ldots,a_n) a_{n+1}.
\end{multline}
Here and below whenever a symbol occurs in the exponential of $(-1)$, it is understood
as its internal degree.

Similarly, let $\cA=(A,Q)$ be a DG-algebra. The differential $Q$ is a degree-1 derivation
$Q:A_p\ra A_{p+1}$ which satisfies $Q^2=0$. Using $Q$, we can make $C^n(A)$ into a complex:
one lets
\begin{multline}
(Qf)(a_1,\ldots,a_n)=Q(f(a_1,\ldots,a_n))\\
-\sum_{i=1}^n (-1)^{v_1+\ldots+v_{i-1}+f+n-1}
f(a_1,\ldots,a_{i-1},Qa_i,a_{i+1},\ldots,a_n).
\end{multline}
Thus on the bigraded vector space $C^*(A)$ we have two differentials: $Q$, which has internal
degree $1$ and $n$-degree $0$, and $\delta$, which has internal degree $0$ and $n$-degree $1$.
The total degree for both differentials is $1$, and it is easy to check that $Q$ and $\delta$
commute. The \Ho\ cohomology of $\cA$ is defined to be the cohomology
of $(-1)^n Q+\delta$. The vector space $\HH^2(\cA)$ classifies infinitesimal deformations of $(A,Q)$
in the category of DG-algebras, up to quasi-isomorphism. More generally, $\HH^*(\cA)$ classifies
deformations of $(A,Q)$ regarded as an $A_\infty$ algebra.

Since \Ho\ cochains are functions taking value in an algebra, the \Ho\ complex has an obvious
algebra structure as well (given by the multiplication of functions). The corresponding
binary product is called the cup product. The \Ho\ coboundary operator $\delta$ is a derivation
of the cup product, and therefore the cup product descends to \Ho\ cohomology, making it into a $\ZZ$-graded algebra. It was noted for the first time by M.~Gerstenhaber~\cite{Ge} that the latter algebra
is always supercommutative, even if $A$ is noncommutative. Gerstenhaber also discovered that there
is a natural graded Lie bracket on $\HH^*(A)$ of degree $-1$. The algebraic structure of $\HH^*(A)$
is formally the same as that of the space of functions on a $\ZZ$-graded supermanifold equipped with a
Poisson bracket of degree $-1$.

There is an equivalent definition of the \Ho\ cohomology of an algebra which has
a nice geometric interpretation. It is this definition that we will generalize in the next
section. Suppose $A$ is commutative; then one can regard $A$ as the algebra of functions on an affine scheme $X=Spec(A)$.
Consider further $A\ot A$, its spectrum, $Spec(A\ot A)=X\times X$, and the diagonal $\Delta\subset X\times X$.
One can think of $\Delta$ as a B-brane (read: object of the bounded derived category)
on $X\times X$, and consider its open-string spectrum (read: endomorphism algebra).
It turns out that the resulting algebra of physical open-string states is precisely the
\Ho\ cohomology of $A$.

An algebraic version of this definition is also very concise. One can think of $A$ as a
left-right bimodule over $A$, or equivalently as a module over $A\ot A$ (here we still
assume that $A$ is commutative). Then the \Ho\ cohomology of $A$ is defined
as
$$
\HH^*(A)=\Ext_{A\ot A}(A,A);
$$
that is, it is the endomorphism algebra of $A$ regarded as an object of the derived
category of modules over $A\ot A$. If $A$ is noncommutative, then $A$ is not a module
over $A\ot A$, but it is a module over $A\ot A^{op}$, where $A^{op}$ is the opposite
algebra of $A$. It turns out that in this more general case we have:
$$
\HH^*(A)=\Ext_{A\ot A^{op}}(A,A).
$$
To see how this comes about, we need to compute the endomorphisms of $\Delta$ in
$D^b(X\times X).$ That is, one has to take a projective
resolution of $A$ regarded as a module over $A\ot A^{op}$, apply to it the operation
$\Hom_{A\ot A^{op}}(-,A)$, and evaluate the cohomology of the resulting complex of vector spaces.
The key
point is that for any algebra $A$ with a unit there is a canonical resolution of $A$
by free $A\ot A^{op}$ modules:
$$
\begin{CD}
\cdots @>>> A^{\ot 4} @>>> A^{\ot 3} @>>> A^{\ot 2}.
\end{CD}
$$
Each term in this complex is a bimodule over $A$, which is the same as a module over $A\ot A^{op}$.
If we use this resolution to compute $\Ext^*(A,A)$, we get the \Ho\ complex.

If we turn our attention to projective or quasi-projective varieties (or schemes)
and their derived
categories of coherent sheaves, several definitions of the \Ho\ cohomology are
possible. One possibility is to sheafify the \Ho\ complex and take the hypercohomology
of the resulting complex of sheaves as the definition of \Ho\ cohomology of $D^b(X)$.
This is the approach taken in Ref.~\cite{WeiGel} for \Ho\ {\it homology}; a similar definition
of \Ho\ cohomology is adopted in Ref.~\cite{GS}. Another possibility is to
consider the diagonal subvariety $\Delta\subset X\times X$ and define the
\Ho\ cohomology of $X$ to be the endomorphism algebra of $\Delta$ regarded as an object of
$D^b(X\times X)$~\cite{Swan}. Remarkably, all these definitions give the same result for quasi-projective schemes~\cite{Swan}. M. Kontsevich interpreted the last definition of $\HH^*(X)$
as computing the space of infinitesimal deformations of the bounded derived
category of coherent sheaves on $X$ in the class of $A_\infty$ categories~\cite{Konts}.

The special thing about the diagonal in $X\times X$
is that it is a geometric object representing the identity functor from the category of D-branes
on $X$ to itself. That is, if one takes a D-brane on $X$, pulls it back to $X\times X$ using the
projection to the first factor, tensors with the diagonal object, and then pushes down to $X$ using the
projection to the second factor, then one gets back the same D-brane, up to isomorphism. Then endomorphisms of the diagonal D-brane parametrize infinitesimal deformations of the
identity functor, i.e. infinitesimal deformations of the category of D-branes on $X$~\cite{Konts}.
Thus it is natural to define the \Ho\ cohomology of the category of D-branes on $X$ as the endomorphism
algebra of the diagonal D-brane on $X\times X$.

Using the ``diagonal brane'' definition, it is easy to see that for smooth quasi-projective
$X$ the \Ho\ cohomology is
\begin{equation}\label{Hon}
\HH^n(X)=\op_{p+q=n} H^p(\Lambda^q TX).
\end{equation}
Indeed, if we are dealing with a B-brane on a manifold $Z$ which corresponds to the structure
sheaf of a smooth submanifold $Y$, then its open string algebra is given by
$$
\op_{p,q}H^p(\Lambda^q NY),
$$
where $NY$ is the normal bundle of $Y$ in $Z$. The fermion number (i.e. grade) is $p+q$.
More precisely, this bigraded vector space
is the first term in a spectral sequence which converges to the space of physical
open strings~\cite{KatzSharpe}.
In our case $Z=X\times X$, and $Y=\Delta$, so $NY=TX$, and the first term of the spectral
sequence computing the space of physical open string states for $\Delta$ is precisely
Eq.~(\ref{Hon}). The spectral sequence actually collapses in this case, so
the \Ho\ cohomology of $X$ is given by Eq.~(\ref{Hon}).

More generally, one notes that instead of the canonical resolution of the diagonal mentioned above
one can take any other projective resolution. This is very useful, because the ``canonical''
resolution is inconvenient for computations: even for $A=\CC[x]$, the algebra of polynomials
in one variable, it has infinite
length. In the next section we will be mostly interested in the case when $X$ is an affine
space $V\simeq\CC^n$. In this case a much more convenient free resolution of $\Delta$
is known, the Koszul resolution. To describe it, let $x_1,\ldots,x_n$ and $y_1,\ldots,y_n$
be affine coordinates on $V\times V$ and let us introduce $n$ fermionic variables
$\theta_1,\ldots,\theta_n$. Consider the vector space $K$ of polynomial functions of all these
variables. It is graded by the negative of the fermion number.
It is easy to see that $K$ is a free graded module over
$\CC[V\times V]$ of rank $2^n$. Consider the following linear map on $K$:
$$
k=\sum_{i=1}^n(x_i-y_i)\frac{\partial}{\partial\theta_i}
$$
Obviously, $k$ is a degree-one endomorphism of $K$ which squares to zero. The pair $(K,k)$
is the Koszul complex for $\Delta$. One can check that its cohomology is concentrated in
degree zero and is isomorphic to $\CC[V\times V]/(x-y)\simeq \CC[V]$.

M. Kontsevich conjectured~\cite{Konts} that the ``diagonal brane'' construction will work in a similar
way for the category of A-branes on a Calabi-Yau $X$. The category of A-branes is the derived
category of the Fukaya category, which is an $A_\infty$ category whose objects are, roughly
speaking, Lagrangian submanifolds with flat vector bundles. The diagonal brane is the
diagonal submanifold in $X\times X$, where the second copy of $X$ is taken with a
symplectic structure
opposite to that on the first copy.\footnote{Apparently, reversing the sign of the symplectic
form is analogous to passing from $A$ to $A^{op}$ for B-branes. This analogy would have
a natural explanation if the Fukaya category were somehow related to the deformation quantization
of $(X,\omega)$: changing the sign of $\omega$ has the effect of replacing the quantized
algebra of functions on $X$ with its opposite. A relation between the Fukaya category and
deformation quantization has been conjectured by many people, including one of the authors
of the present paper; unfortunately, the nature of such a relation remains elusive.}
We will denote the second copy of $X$ with a reversed symplectic structure by $X^{op}$.
The vector bundle on $\Delta\subset X\times X^{op}$ is taken to be the trivial rank one bundle. Kontsevich
conjectured that the endomorphism
algebra of this diagonal brane, regarded as an object of the derived Fukaya category, is
isomorphic to the quantum cohomology ring of $X$ (which is the algebra of physical closed string states in the A-model on $X$). This conjecture remains unproved.

One can give a general (although somewhat nonrigorous) argument that the \Ho\ cohomology of the
category of D-branes defined as the endomorphism algebra of the diagonal brane is the space of physical
closed string states.
Consider a closed-string spherical worldsheet $\Sigma$ on which there lives some topological
field theory $\cT$. We can reinterpret it as an open-string disk worldsheet as follows. Let us draw an equatorial circle (seam) on $\Sigma$ and identify the upper and lower hemispheres by means of an antiholomorphic map. Let $\cT^{op}$ be the image of $\cT$ under this map. (For B-model on a
Calabi-Yau, $\cT^{op}$ is isomorphic to $\cT$; for a Landau-Ginzburg model, $\cT^{op}$ differs
from $\cT$ only by the reversal of the sign of the superpotential; for A-model on a Calabi-Yau, $\cT^{op}$
differs from $\cT$ by the reversal of the sign of the symplectic structure.) Then $\cT$
living on $\Sigma$ is equivalent to $\cT\ot\cT^{op}$ living on the upper hemisphere, with some gluing
condition on the boundary. This gluing condition corresponds to a D-brane on $X\times X^{op}$, and it is
chosen so that after reinterpreting a $\cT\ot\cT^{op}$ field configuration on the upper hemisphere
as a $\cT$ field configuration on the union of upper and lower hemispheres, the fields join
smoothly on the boundary. This is a physical definition of the diagonal brane. If a single closed-string
operator is inserted on $\Sigma$, we may choose the seam to pass through the insertion point, and then
we should be able to interpret it as an open-string operator in $\cT\ot\cT^{op}$ inserted on the
boundary of the disk. Continuing this line of argument, we see that the algebra of physical closed-string
states should be isomorphic to the endomorphism algebra of the diagonal brane, and that this
isomorphism should identify the closed-string topological metric with the open-string
topological metric.\footnote{Recall that for any algebra or DG-algebra $A$
its \Ho\ cohomology has the structure of a supercommutative algebra~\cite{Ge}.
This is in agreement with the fact that
the algebra of physical closed-string states is always supercommutative.}
In this paper we check these statements in the case of topological Landau-Ginzburg models on
the affine space and on certain orbifolds of the affine space.

\section{Landau-Ginzburg models on affine space}

Consider a topological Landau-Ginzburg model on the affine space $V\simeq \CC^n$ with a polynomial
superpotential $W:V\ra \CC$.\footnote{In what follows we will always assume that the critical set of $W$
is compact; this is necessary for the Landau-Ginzburg field theory to have a normalizable vacuum state.}
To this data one can associate a CDG-algebra, i.e. a $\ZZ_2$-graded
algebra with an odd derivation $Q:A\ra A$ and an even element $B\in A$ such that $Q^2a=[B,a]$ for
any $a\in A$. The derivation $Q$ is sometimes called a twisted differential
(ordinary differentials satisfy $Q^2=0$).
The notion of a CDG-algebra is a slight generalization of the notion of a ($\ZZ_2$-graded)
DG-algebra. In the Landau-Ginzburg case, $A=\CC[V]$ (the algebra of polynomials on $V$), $Q=0$,
and $B=W$. A CDG-module over a CDG-algebra $(A,Q,B)$ is a pair $(M,D)$, where $M$
is a graded module over $A$ and $D$ is an odd derivation of $M$ such that $D^2m=Bm$ for any
element $m\in M$. In the Landau-Ginzburg case, $Q=0$, so $D$ is simply an odd endomorphism
of $M$ satisfying $D^2=W$.

As explained in Refs.~\cite{KLi1,KLi2}, the category of D-branes for the topological LG model
is equivalent to a DG-category whose objects are free CDG-modules over the Landau-Ginzburg CDG-algebra,
and morphisms are morphisms between CDG modules. That is, the space
of morphisms between $(M,D)$ and $(M',D')$ is a differential graded vector space, which
coincides with the space of morphisms between graded modules $M$ and $M'$, as far as graded
vector structure is concerned, while the differential is
$$
D_{MM'}(\phi)=D'\circ\phi - (-1)^{|\phi|} \phi\circ D,\quad \phi\in \Hom_A(M,M').
$$
One can make an ordinary category out of this DG-category by redefining morphisms between $M$ and $M'$
to be the cohomology of $D_{MM'}$. This is the homotopy category of CDG-modules. In the affine case,
this is the same as the derived category of CDG-modules.
The space of physical open-string states is identified with the cohomology of $D_{MM'}$. In particular,
if we take $M=M'$, then the algebra of physical open-string states is identified with the
endomorphism algebra of $(M,D)$ in the derived category of CDG-modules.

If $M$ is a free $\ZZ_2$-graded module over $A=\CC[V]$, then the endomorphism algebra of $M$
(in the category of graded $A$-modules) is simply the algebra of matrices with entries
in $\CC[V]$. This algebra is graded in an obvious way: endomorphisms which preserve the parity
of elements of $M$ are considered even, while endomorphisms which flip the parity are
considered odd. If we order the generators of $M$ so that even generators go first, then
elements of $\End_A(M)$ can be written as $2\times 2$ block matrices with entries in $A$,
where the diagonal blocks are even, and the off-diagonal ones are odd. Such matrices are usually called
supermatrices (over $A=\CC[V]$). The problem of finding solutions to the equation $D^2=W$ can be thought
of as the problem of factorizing $W$ into a product of two identical odd supermatrices
with polynomial entries. We will sometimes refer to it as the matrix factorization problem.

The goal of this section is to compute the \Ho\ cohomology of the category of D-branes
in topological LG models and to compare it
with the algebra of closed string states, which is known to be the so-called Jacobi ring
of $W$, i.e.
\begin{equation}\label{milnor}
\JRW=\CC[V]/\IDW,\quad \IDW=(\partial_1 W,\ldots,\partial_n W).
\end{equation}
But first we need to give a definition of the \Ho\ cohomology of a CDG-algebra. By analogy
with the previous section, we would like to define it as the endomorphism algebra of
the ``diagonal'' brane on $V\times V$, or more algebraically, as the endomorphism algebra of a
$A=\CC[V]$ regarded as an object of the derived category of CDG-bimodules over itself.
That is, we would like to regard $A$ as a CDG-module over the tensor product of $(A,0,W)$ with
its opposite. Since $A$ is commutative, one may ask why we are retaining the word
``opposite'' here.
We claim that in the world of CDG algebras taking an opposite of an algebra includes changing
the sign of the special element $B$. Indeed, suppose we take some arbitrary CDG-algebra
$(A,Q,B)$. We cannot regard it as a CDG-module over itself by simply setting $D=Q$, because
the defining relation for a CDG-module is $D^2(a)=B a$, while $Q$ satisfies $Q^2a=[B,a]$.
Next, if we consider the tensor product of $(A,Q,B)$ with $(A^{op},Q,B)$, we still
cannot regard $(A,Q,B)$ as a CDG-module over this tensor product algebra, because
setting $D=Q$ still gives $D^2(a)=[B,a]$, while the defining relation of the CDG-module
gives in this case
$$
D^2(a)=Ba+aB.
$$
However, if we take the tensor product of $(A,Q,B)$ with $(A^{op},Q,-B)$, then setting
$D=Q$ does make $(A,Q)$ into a CDG-module over the tensor product algebra.

In the Landau-Ginzburg case, this means that we define \Ho\ cohomology as the endomorphism
algebra of $(\CC[V],0,W)$ regarded as a B-brane on $V\times V$ with the superpotential $W(x)-W(y)$,
where $x,y$ are affine coordinates on the two copies of $V$. To compute this endomorphism
algebra, we have to replace $(\CC[V],0,W)$ by a free CDG-module which is isomorphic
to $(\CC[V],0,W)$ in the sense of the derived category of CDG-modules.

We can simply guess the equivalent free CDG-module (a more rigorous justification
is given in the end of this section). If $W$ were
zero, then we would have to replace the structure sheaf of the diagonal with its free
resolution. Since we are dealing with CDG-modules, we have to find instead a ``twisted''
resolution, i.e. a sequence of free $\CC[V\times V]$-modules and morphisms between them such that
the composition of successive morphisms is $W$ instead of zero. Therefore, we will do the
following. We will take the Koszul resolution of the diagonal $\Delta\subset V\times V$, fold
it modulo $2$ (since our algebra is $\ZZ_2$-graded, complexes must also be $\ZZ_2$-graded),
and then deform in a natural way the differential into a twisted differential.
This will give us a free CDG-module on $V\times V$, and we will take its endomorphism
algebra as the definition of the \Ho\ cohomology for the category of D-branes.

First let us see how this works for the one-variable case, where $V\sim \CC$, and $A=\CC[x]$.
We are considering the diagonal submanifold $\Delta$ in $\CC^2$ given by the equation $x=y$.
The Koszul resolution is very short:
$$
\begin{CD}
\CC[x,y]@>{x-y}>>\CC[x,y]
\end{CD}
$$
The folded resolution can be regarded as a $\ZZ_2$-graded module of the form
$\CC[x,y]\op\CC[x,y]$ and the differential
$$
D=\begin{pmatrix} 0 & 0 \\ x-y & 0\end{pmatrix}
$$
This is an honest differential satisfying $D^2=0$. On the other hand, a twisted differential
would satisfy $D^2=W(x)-W(y)$. The correct deformation is obvious:
$$
D=\begin{pmatrix} 0 & \frac{W(x)-W(y)}{x-y} \\ x-y & 0\end{pmatrix}
$$
Note that since $W(x)$ and $W(y)$ are polynomials, $D$ has polynomial entries.
This matrix factorization has been previously considered in a similar context in Ref.~\cite{KhRo}.

Above $D$ was written as an odd supermatrix. Equivalently,
we can think about such supermatrices as differential operators on functions of odd variables.
(This is the approach we took in the last section to write down the Koszul resolution).
In the fermionic language the deformed $D$ looks as
follows:
$$
D=(x-y)\frac{\partial}{\partial\theta}+\frac{W(x)-W(y)}{x-y}\,\theta.
$$

According to Ref.~\cite{KLi1,KLi2,Letal}, the endomorphism algebra of this brane is the
cohomology of the following linear operator on the space of supermatrices
with polynomial entries:
$$
\cD:\phi\ra D\phi-(-1)^{|\phi|}\phi D,
$$
where $\phi$ is a supermatrix. For the operator $D$ as above, it is easy to
see that the cohomology is purely even and isomorphic to the quotient vector space
$$
\CC[x,y]/I,\quad I=\left(x-y,\frac{W(x)-W(y)}{x-y}\right)
$$
Obviously, this vector space is isomorphic to $\JRW=\CC[x]/(\partial_x W(x))$, as claimed.
In fact, it is easy to see that the isomorphism holds on the level of algebras, rather
than on the level of vector spaces.

Next consider the case when $V\simeq \CC^n$. Let us introduce
convenient condensed notations:
\begin{equation}\label{l1}
\bx = (x_1,\ldots,x_n),\qquad \byxi =
(y_1,\ldots,y_i,x_{i+1},\ldots,x_n)
\end{equation}
and for a polynomial $\xP\in\CC[\bx]$ define
\begin{equation}\label{l2}
\DliP = \frac{\xP(\byximo) - \xP(\byxi)}{x_i - y_i}\in\CC[\bx,\by].
\end{equation}

To write the correct free CDG-module
(the twisted resolution of the diagonal), we first write $W(\bx)-W(\by)$ as a sum of polynomials
\begin{equation}\label{sumW}
W(\bx)-W(\by)=\sum_{i=1}^n (x_i-y_i)\Delta_i(W).
\end{equation}
%where
%\begin{equation}
%\Delta_i(W)=W(y_1,\ldots,y_{i-1},x_i,\ldots,x_n)-
%W(y_1,\ldots,y_i,x_{i+1},\ldots,x_n).
%\end{equation}
We want to write $W(\bx)-W(\by)$ as $D^2$, where $D$ is a odd endomorphism of free $\ZZ_2$-graded
modules. The key point is that for each term $\Delta_i(W)$ we can accomplish this by mimicking
the one-variable case. That is, if we let
\begin{equation}\label{Di}
D_i=(x_i-y_i)\frac{\partial}{\partial\theta_i}+\Delta_i(W)\,\theta_i,
\end{equation}
then $D_i$ is a odd endomorphism of free graded $\CC[x,y]$-modules
which satisfies $D_i^2=\Delta_i(W)$.
Therefore if we let
\begin{equation}\label{diagonaln}
D=\sum_{i=1}^n D_i,
\end{equation}
we will get the desired $D^2=W(\bx)-W(\by)$. Moreover, if we take the limit $W\ra 0$, the
differential $D$ will reduce to $k$, the Koszul differential from the previous section.
The corresponding matrix factorization has been previously discussed in Ref.~\cite{KhRo}.

Since the differential is given by a sum, and we have computed the cohomology of $D_i$
already, the cohomology of $D$ is simply
$$
\CC[V\times V]/I,\quad I=\big(x_1-y_1,\ldots,x_n-y_n,\Delta_1(W),\ldots,
\Delta_n(W)\big),
$$
Obviously, this is isomorphic to Eq.~(\ref{milnor}).

We end this section by explaining why the D-brane
Eq.~(\ref{diagonaln}) deserves to be called the diagonal brane. The general definition of the diagonal
is that it represents the identity functor in the category of D-branes on $(\CC^n,W)$.
In the present case, this means the following. Let $(R,D)$ be the matrix factorization of
$W(\bx)-W(\by)$ corresponding to Eq.~(\ref{diagonaln}). Thus $R$ is a free graded $A\ot A$-module,
where $A\simeq \CC[\bx]$, and $D$ is an odd endomorphism of $R$ satisfying $D^2=W(\bx)-W(\by)$.
Let $(N,F)$ be an arbitrary matrix factorization of
$W(\by)$ representing some D-brane on $(\CC^n,W)$. Thus $N$ is a free graded $A$-module, and
$F$ is an odd endomorphism of $N$ satisfying $F^2=W(\by)$. Consider the graded tensor product $(R\ot_A N,D+F)$.
The module $R\ot_A N$ is a free $A$-module, and $D+F$ satisfies
$$
(D+F)^2=D^2+F^2=(W(\bx)-W(\by))+W(\by)=W(\bx).
$$
Thus the pair $(R\ot_A N,D+F)$ is a D-brane on $(\CC^n,W)$ (of infinite rank). By definition,
$(R,D)$ is the diagonal iff $(R\ot_A N,D+F)$ is isomorphic to the original
finite-dimensional brane $(N,F)$, for any $(N,F)$ (in the homotopy category of CDG-modules).
This statement was proved in Ref.~\cite{KhRo} (Prop.~23 of that paper).

\section{LG orbifolds}

\subsection{Minimal model I}

Consider the LG model with target $\CC/\ZZ_n$:
$$
W_0=x^n,\quad x\sim \zeta x,\ \zeta=e^{\frac{2\pi i}{n}}
$$
According to Ref.~\cite{IV}, the chiral ring is simply the invariant part of the Jacobi
ring, which is $\CC[x]/x^{n-1}$. In other words, there are no chiral primaries in the twisted
sectors. The only invariant is actually the identity element.
Let us check this using the diagonal B-brane approach.

We consider a matrix factorization of the potential
$$
W=x^n-y^n.
$$
$W$ is a function on $\CC^2$ invariant under a $\ZZ_n\times \ZZ_n$ action
$$
x\ra \zeta^k x,\quad y\ra \zeta^{k+l} y,\quad k,l\in \ZZ/n.
$$
In other words, $x$ has weight $(1,0),$ while $y$ has weight $(1,1)$.
The corresponding twisted differential $D$ is
$$
D_0(x,y)=\begin{pmatrix} 0 & \prod_{i=1}^{n-1} (x-\zeta^i y)\\ x-y & 0\end{pmatrix}
$$
$D_0$ is regarded as an odd endomorphism of a free $\ZZ_2$-graded module over $\CC[x,y]$ of rank two,
which we write as $M_0=M_0^+\op M_0^-$, where $M_0^+\simeq\CC[x,y]$ is the even component, and
$M_0^-\simeq \CC[x,y]$ is the odd component.

We need to define a $G=\ZZ_n\times\ZZ_n$ action on $M_0$ so that $M_0$ becomes an equivariant
module over $\CC[x,y]$, and the twisted differential
is an equivariant endomorphism. A $\ZZ_n$ action on a free module, like $M_0^+$ or $M_0^-$, is completely specified once we specify the action on the unit element, i.e. its weight. For the first
$\ZZ_n$, we choose the weights to be $0$ for $M_0^+$ and $1$ for $M_0^-$. This is the only way to make
$D_0$ equivariant, up to an overall shift of weights. For the second $\ZZ_n$, no choice of
weights works, so we have to equivariantize $D_0$, i.e. to add ``image branes'':
$$
D_l(x,y)=D_0(x,\zeta^l y)=\begin{pmatrix} 0 & \prod_{i=0,i\neq l}^{n-1} (x-\zeta^i y)\\
x-\zeta^l y & 0 \end{pmatrix},\quad l=1,\ldots,n-1.
$$
$D_l$ acts on a module $M_l$ which is isomorphic to $M_0$ as a $\ZZ_2\times \ZZ_n$-graded module.
We take the diagonal brane $M$ to be the direct sum of all these branes, with the twisted
differential $D$ being the sum of all $D_l$, $l=0,\ldots,n-1$. $D$ is equivariant with respect to
the second $\ZZ_n$ if we let $\ZZ_n$ act on the summands $M_l$ by cyclic permutations.

Now let us compute the endomorphism algebra of the diagonal brane $(M,D)$. First we compute the
endomorphisms disregarding the $\ZZ_n\times\ZZ_n$ action, and then project onto the invariant
part. Since $M$ is a direct sum, we can compute morphisms between the summands,
and then sum them up. It is also convenient to consider separately the morphisms from a brane
$(M_l,D_l)$ to itself, and to its mirror images $(M_k,D_k)$, $k\neq l$. An easy computation shows that
the endomorphisms of $(M_l,D_l)$ (i.e. the cohomology of $D_l$ in the adjoint representation) is
purely even and is the Jacobi ring of $W_0$. The space of morphisms between $(M_l,D_l)$ and $(M_k,D_k)$ is
purely odd and one-dimensional. A basis element of this odd vector space will be called
$\theta_{l,k}$. Since the odd components of $M_l$ have weight $1$ with respect to the first
$\ZZ_n$, one can easily see that $\theta_{k,l}$ has weight $1$. Projecting with respect to the
first $\ZZ_n$, we see that no morphisms between a brane and its image survive. As for
endomorphisms of $(M_l,D_l)$, only the invariant part of the Jacobi ring (which consists only of the
subspace spanned by the identity) survives. Thus we get the direct sum of $n$ copies of
the trivial algebra. Finally, we project with respect to the second
$\ZZ_n$. This $\ZZ_n$ cyclically permutes all $M_l$, and therefore identifies all the $n$ copies
of the trivial algebra.

\subsection{Minimal model II}

This is a generalization of the previous subsection. We consider a LG model with target
$\CC/\ZZ_n$ and a superpotential
$$
W_0=x^{nm},\quad n,m\in\NN.
$$
The orbifold group acts by
$$
x\mapsto \zeta^k x, \quad k\in \ZZ/n.
$$
The analysis is very similar to the previous subsection. The untwisted sector is the invariant
part of the Jacobi ring. There are still $n-1$ image branes, and morphisms between different images
are odd and have weight $1$ under the first $\ZZ_n$ action. Hence the twisted sector is killed
by the first $\ZZ_n$ projection. This agrees with the results of Ref.~\cite{IV}.

\subsection{Gepner model for a Calabi-Yau 0-fold}

This is a LG orbifold with a target $\CC^2/\ZZ_2$ and superpotential
$$
W_0=x_1^2+x_2^2.
$$
The $\ZZ_2$ action is diagonal (flips the sign of both variables).
E.~Witten's argument~\cite{phases} shows that this is a Gepner model for the CY 0-fold
given as a hypersurface $x_1^2+x_2^2=0$ in $\PP^1$. The latter consists of two points,
so the corresponding chiral ring is two-dimensional (the direct sum of two copies of the
trivial algebra). Let us check this using the diagonal brane approach.

We consider the superpotential
$$
W=x_1^2-y_1^2+x_2^2-y_2^2=W_1(x_1,y_1)+W_2(x_2,y_2).
$$
To construct the diagonal brane, we factorize separately $W_1$ and $W_2$ and then take the
$\ZZ_2$-graded tensor product
of the corresponding branes. If we represent $D_1$ and $D_2$ by fermionic differential
operators, this simply means that we let $D=D_1 + D_2$, where $D_1^2=W_1$ and $D_2^2=W_2$.
Since $D_1$ and $D_2$ anticommute, this gives $D^2=W$.

The factorizations of $W_1$ and $W_2$ are taken as in the previous subsection. Then we have to
equivariantize with respect to $\ZZ_2\times\ZZ_2$. As above, equivariance with respect to
the first $\ZZ_2$ forces the odd component of $M$ to have weight $1$, if the even component
has weight $0$. Hence odd endomorphisms of $M$ have weight one, while even endomorphisms have
weight $0$. Equivariance with respect to the second $\ZZ_2$ requires adding a single ``image brane''
$(M',D')$, where $M'\simeq M$, and $D'$ is
$$
D'(x_1,x_2,y_1,y_2)=D(x_1,x_2,-y_1,-y_2).
$$
The diagonal brane is the direct sum of $(M,D)$ and its image $(M',D')$. Let us compute its
endomorphism algebra
and then project onto the $\ZZ_2\times\ZZ_2$-invariant part. First consider the endomorphisms
of $M$ and $M'$. Since both $M$ and $M'$ have factorized form (they are graded tensor products
of $M_1$ and $M_2$), we can compute the endomorphisms of $M_1$ and $M_2$ and then tensor them.
{}From the previous subsection we known that the endomorphism algebras of $M_1$ and $M_2$ are trivial (spanned by the identity). Hence the endomorphism algebras of $M$ and $M'$ are also trivial.
They also survive the first $\ZZ_2$ projection, while the second $\ZZ_2$ projection identifies
them.

Now consider morphisms from $M$ to $M'$. Using the results of the previous subsection, we see
that the space of morphisms is one-dimensional and spanned by
$\theta_{0,1}^{(1)}\ot \theta_{0,1}^{(2)}$,
where $\theta_{0,1}^{(i)}$ spans the space of morphisms from $M_i$ to its image under the
second $\ZZ_2$. Thus the space of morphisms from $M$ to $M'$ is purely even and one-dimensional.
Since both $\theta$'s have weight $1$ under the first $\ZZ_2$, their product has weight $0$
and survives the first $\ZZ_2$ projection. Finally, the second $\ZZ_2$ projection identifies
morphisms from $M$ to $M'$ and from $M'$ to $M$. Thus the twisted sector contributes a single
even element to the chiral ring. It is easy to see that this element squares to identity.
Thus the endomorphism algebra of the diagonal brane isomorphic to the group ring of $\ZZ_2$.
This is the desired result.

\subsection{Gepner model for a Calabi-Yau $n-2$-fold}

We consider the superpotential $W=x_1^n+\ldots + x_n^n$ on $\CC^n$ with a diagonal $\ZZ_n$ action:
$$
x_i\mapsto \zeta x_i.
$$
This is a Gepner model for the Fermat hypersurface $x_1^n+\ldots+x_n^n=0$ in $\PP^{n-1}$.
Let us compute the endomorphisms of the diagonal brane. The computation proceeds along the
same lines as in the previous subsection. We now have $n-1$ image branes (with respect to the
second $\ZZ_n$). The untwisted sector gives us the invariant part of the Jacobi ring.
Morphisms between a brane and its image have the form
$$
\ot_{i=0}^{n-1} \theta_{k,l}^{(i)}
$$
Each $\theta$ is odd and has weight $1$ under the first $\ZZ_n$, so the product has weight $0$ and survives the first $\ZZ_n$ projection. It is even or odd depending on whether $n$ is even or odd.
The second $\ZZ_n$ projection identifies all the twisted sector states with the same value of
$k-l$. Hence the twisted sector contributes $n-1$ states to the chiral ring
(labeled by $k-l\ {\rm mod}\ n$).

Let us compare this with the B-model of the Fermat hypersurfaces. The case $n=2$ has been discussed
in the previous subsection. For $n=3$ the Fermat hypersurface is a cubic curve in $\PP^2$, i.e.
an elliptic curve. Its B-model is isomorphic to the exterior algebra with two odd generators, i.e.
it has two-dimensional even subspace and two-dimensional odd subspace. On the LG side, the untwisted
sector is spanned by $1,x_1 x_2 x_3$ (the invariant part of the Jacobi ring), while the
twisted sector is odd and two-dimensional. Thus the two agree, as $\ZZ_2$-graded vector spaces.
One can check that the ring structure also agrees.

For $n=4$ the Fermat hypersurface is a quartic in $\PP^3$, i.e. a K3 surface. The B-model
is a purely even algebra of dimension 24. On the other hand, the $\ZZ_4$-invariant part of the Jacobi
ring is easily seen to have dimension 21. The twisted sector gives the other 3 states.

For $n=5$ we are dealing with a quintic in $\PP^4$. The B-model has both an even and odd components.
The even component has dimension 204. The odd component has dimension 4. It is easy to show that
the invariant part of the Jacobi ring is indeed 204-dimensional. The odd component comes entirely
from the twisted sector.

\section{Open-closed correlators for some Landau-Ginzburg models}

\subsection{Closed topological metric from the open one}

In this subsection we discuss how to deduce the 3-point topological
closed-string correlators (the Yukawa couplings) in the diagonal
brane approach and show that the result agrees with the conventional
approach~\cite{Vafa}.

The closed strings form a supercommutative \Frob\ algebra $\xAcl$: there
is a trace map $\xAcl\xrightarrow{\trF}\CC$, which defines the
scalar product on $\xAcl$ and determines the 3-point correlator of
closed string operators
\begin{equation}\label{l7}
\scpr{a}{b}=\trF(ab),\quad\corr{abc}=\trF(abc).
\end{equation}
%
%by the formula $$\scpr{a}{b}=\trF(ab).$$
%\Frob\ trace also determines the 3-point function for closed
%strings:
%
%\begin{equation}\label{l7}
%\corr{abc} = \trF(abc).
%\end{equation}

The algebra of open-string states $\xAop$ has the $\xAcl$-linear `boundary trace' map
$\xAop\xrightarrow{\trb}\xAcl$, which expresses the boundary state
corresponding to a boundary with an open string operator
insertion. The composition of the boundary and \Frob\ traces
$\Tr=\trF\trb$
determines the scalar product and the 3-point correlator for open
strings:
\begin{equation}\label{l8}
\scpr{A}{B}=\trb(AB),\quad\corr{ABC}=\trb(ABC).
\end{equation}

The conjectured correspondence between the diagonal D-brane states
and the closed string states implies that the \Frob\ trace on
$\xAcl$ should be equal to the combined trace $\Tr$ on the
diagonal D-brane states. This would also imply the equality
between their scalar products and 3-point correlators.

We will verify the equivalence of \Frob\ and combined traces for a topological LG
model with the superpotential $W(\bx)\in\CC[\bx]$. In this case the algebra of
closed strings is the Jacobi algebra $\JRW$ of Eq.~\rx{milnor} and the
\Frob\ trace $\trx$ was determined by C.~Vafa in \rf{Vafa}:
\begin{equation}\label{l9}
\trx(\xp)=\frac{1}{(2\pi i)^n}\oint \frac{\xp\ dx_1\wedge \ldots \wedge dx_n}{\partial_1W\ldots
\partial_nW},\quad p\in\CC[\bx].
\end{equation}
The integrand in this formula is a meromorphic $n$-form, and the integration goes over a Lagrangian $n$-cycle
defined by the equations
$$
|\partial_i W|=\eps_i>0, \quad i=1,\ldots,n.
$$

The boundary trace formula for a B-brane in a LG model, presented by a matrix factorization, was
derived in \rf{KLi2} (see also \rf{HerbstLaz}):
\begin{equation}\label{l10}
\trb(\xO) = (1/n!)\; \str \big( \Dwn \xO \big),
\end{equation}
where $\xO$ is an endomorphism regarded as a supermatrix with
polynomial entries, $\str$ is the supertrace, and
\begin{equation}\label{l11}
\Dwn = \ssSnmos\,\partial_{\sgv{1}}D\cdots\partial_{\sgv{n}}D,
\end{equation}
where $\Sn$ is the symmetric group of $n$ elements.

Since the space of endomorphisms of the diagonal B-brane is
isomorphic (up to the BRST equivalence) to the Jacobi ring $\JRWx$,
to prove the equivalence between closed- and open-string traces it is
sufficient to show that
\begin{equation}\label{l12}
\trF\trb(f(\bx)\Id) = \trx(f).
\end{equation}
Here $f(\bx)$ is any polynomial, and $\Id$ is the identity endomorphism of
the diagonal B-brane.

The relation\rx{l12} has an interesting algebraic corollary. Let
$\Acn$ be the dual space of an algebra $A$, then the
multiplication map $A\xrightarrow{m}A\otimes A$ has a dual map
$\Acn\otimes\Acn\xrightarrow{\Delta} \Acn$, which is called
\emph{\cmult}. The scalar product\rx{l7} establishes a canonical isomorphism
between the \Frob\ algebra $\xAcl=\CC[\bx]/I_W$ and its dual, so there is a
\cmult\ map
\begin{equation}\label{l1.1}
\xAcl\xrightarrow{\Delta}\xAcl\otimes\xAcl
\end{equation}
with the property
\begin{equation}\label{l1.2}
\Delta(a) = (a\otimes 1)\,\Delta(1) = (1\otimes a)\,\Delta(1).
\end{equation}
The relation\rx{l12} implies that
\begin{equation}\label{l1.3}
\trb(\Id) = \Delta(1)\in (\CC[\bx]/I_W)\otimes(\CC[\by]/I_W).
\end{equation}
%

%Since the space of endomorphisms of the diagonal B-brane is
%isomorphic (up to the BRST equivalence) to the Jacobi ring $\JRWy$,
%then the equivalence between the closed- and open-string traces
%would follow from the following relation
%
%\begin{equation}\label{l12}
%\trx\trb(\Id) = 1,
%\end{equation}
%
%where $\Id$ is the identity endomorphism of the diagonal B-brane.
%In fact, this relation (together with the fact that $(x_i-y_i)\trb(Id) =0$
%as an element of the $(\bx,\by)$ Jacobi ring $\JRWx\otimes\JRWy$)
%would also imply an important algebraic property of $\trb(Id)$:
%$$\trb(\Id) = \Delta(1),$$
%that is $\trb(\Id)$ is the comultiplication of the unit element of
%$\JRWx$ (comultiplication is the dual operation to the
%multiplication; the duality is based on the scalar product\rx{l2}
%of the Jacobi ring as a \Frob\ algebra).

As the first step of proving the relation\rx{l12}, we are going to
derive a convenient formula for $\trb(\Id)$:
\begin{equation}\label{l3}
%\frac{1}{(2n)!}\;
\trb(\Id) = \frac{1}{(2n)!}\;\strD =
\det \xmtr{\DlijW}_{i,j=1}^n
%\ssSnmos \prion\DlidW,
\end{equation}
where
%$\Sn$ is the symmetric group of $n$ elements and
$\partial_j W(\bx) = \partial W(\bx)/\partial x_j$
and $\Delta_i$ is defined by \ex{l2}.
Indeed, if we expand the $2n$-th power of the sum\rx{Di}, then the
supertrace of an individual monomial is zero, unless it contains
all $\theta$'s and all $\partial/\partial\theta$'s. Therefore, if
we lump all variables together $\bfz = (\bx,\by)$, then
\begin{equation}\label{l4}
\strD = \frac{(2n)!}{2^n}\ssStnmos\prion
\xCzitio,
\end{equation}
where
\begin{equation}\label{l5}
\xCzijk =
\str_i
\left(\prtzDistj\; \prtzDistk\right),
%(\partial_{\bfz_{\sgti}} D_i)\,(\partial_{\bfz_{\sgtio}},D_i),
\end{equation}
while $D_i$ is defined by \ex{Di} and $\str_i$ denotes the trace
over the space $(1,\theta_i)$. A direct computation shows that if
$\{z_j,z_k\}\cap\{x_i,y_i\}=\emptyset$, then $\xCzijk=0$ and the
only non-zero factors are
\begin{eqnarray}\label{l6}
%\xCiv{x_i}{y_i} = -\xCiv{y_i}{x_i} = \Dliv{\partial_i W},
%\\
\xCiv{x_j}{y_i} = -\xCiv{y_i}{x_j} = \Dliv{\partial_j
W},\qquad j\leq i,
\\
\xCiv{x_i}{y_j} = -\xCiv{y_j}{x_i} = \Dliv{\partial_j W},\qquad
j\geq i.
\end{eqnarray}
Thus, a non-zero supertrace $\str_i$ must contain at least one
derivative over an $i$-th variable ($x$ or $y$). The distributions of
derivatives over the supertraces, which do not produce zero factors, can be enumerated by the elements
$\sg\in\Sn$: if $\sg(i)=i$, then the supertrace\rx{l5} includes
the derivatives over $x_i$ and $y_i$; if $\sg(i)\neq i$, then one
derivative of\rx{l5} should be over the $\sg(i)$-th element of the list
$\byxi$ and the other should be either over $x_i$ or over $y_i$
depending on whether $\sgin(i)>i$ or $\sgin(i)<i$.
This combinatorics together with the expressions\rx{l6} leads to the
formula\rx{l3}.

It remains to apply the Frobenius trace to the r.h.s. of Eq.~(\ref{l3}) multiplied by $f(\bx)$.
To do this, it is convenient to assume that all critical points of $W$ are non-degenerate.
Since the set of such superpotentials is an open and dense subset of all admissible
superpotentials, and the l.h.s. of Eq.~(\ref{l12}) is a continuous function of the coefficients
of $W$, it is sufficient to prove Eq.~(\ref{l12}) in this special case. For $W$ with only
non-degenerate critical points, the trace function Eq.~(\ref{l9}) becomes
\begin{equation}\label{l2.2}
\trF(\xp)=\sum_{\rm crit\ W} \frac{p(\bx)}{{\rm Hess}\, W(\bx)},
\end{equation}
where ${\rm Hess}\, W$ denotes the Hessian of $W$, and the sum is over all critical points of $W$.
Thus we simply have to evaluate the r.h.s. of Eq.~(\ref{l3}) at the critical points of $W(\bx)-W(\by)$,
which are pairs of critical points of $W(\bx)$.

Let $\bx$ and $\by$ be two critical points of $W$. If
$\bx\neq\by$, then $\trb(\Id)=0$ in view of \ex{l3}\footnote{We
are thankful to Christiaan Hofman for pointing out an error in our original proof and for suggesting
the correct one.}. Indeed, let $$r_i = \xrw{\DlijW}_{j=1}^n$$ be
the rows of the matrix of \ex{l3}. Obvious cancellations
lead to the following identity:
\begin{equation}\label{l2.1}
\sum_{i=1}^n (x_i - y_i)\,r_i = \xrw{\partial_j W(\bx) - \partial_j
W(\by)}_{j=1}^n.
\end{equation}
Since $\bx$ and $\by$ are critical points of $W$, the \rhs\
of this equation is zero, and if $\bx\neq\by$, then this means
that the rows $r_i$ are linearly dependent, and the determinant in
\ex{l3} is equal to zero.

Thus, only the critical points on the diagonal, i.e. the ones for which $\bx=\by$,
contribute to the sum\rx{l2.2}.
Expanding $W$ into a Taylor series to quadratic order, one
can easily see that the r.h.s. of Eq.~(\ref{l3}) evaluated on such a critical point is equal to
${\rm Hess}\, W$. Hence we get:
\begin{equation}
\trF\trb(f(\bx)\Id)=\sum_{\rm crit\ W} \frac{f(\bx)}{{\rm Hess}\, W(\bx)}=\trx(f).
\end{equation}
This completes the proof of Eq.~(\ref{l12}).

As another example, let us deduce the topological closed-string metric
for the elliptic curve at the Fermat point, using the LG representation and the
diagonal brane approach. First, let us state the expected answer. The closed-string
algebra of an elliptic curve is isomorphic to the exterior algebra with two generators
$\xi_1,\xi_2$. Thus the space of physical closed-string states is four-dimensional,
with two-dimensional even subspace (the sum of $H^0(\O_X)$ and $H^1(T_X)$)
and two-dimensional odd subspace (the sum of $H^0(T_X)$ and $H^1(\O_X)$).
The trace on this algebra is nonvanishing only on the one-dimensional subspace
spanned by $\xi_1\xi_2$, which geometrically corresponds to $H^1(T_X)$.

In the LG approach, the states corresponding to $\xi_1$ and $\xi_2$
come from the twisted sector, while $1$ and $\xi_1\xi_2$ span the
$\ZZ_3$-invariant part of the Jacobi ring of the superpotential
$$
W=x_1^3+x_2^3+x_3^3.
$$
If we think about the closed-string space of states as the endomorphism
algebra of a diagonal brane on $(\CC^3/\ZZ_3)\times(\CC^3/\ZZ_3)$,
then the endomorphisms corresponding to $\xi_1$ and $\xi_2$ are off-diagonal
(they correspond to strings from a brane to one of its images under a $\ZZ_3$),
and it follows from Eq.~(\ref{l10}) that their trace vanishes. As for the
endomorphisms corresponding to $1$ and $\xi_1\xi_2$, they are insensitive to the
orbifolding and their properties are exactly the same as in the LG model
on the affine space. We already know that the open-string trace for these
endomorphisms coincides with the closed-string trace Eq.~(\ref{l9}),
provided we correctly identify endomorphisms with the elements of the Jacobi
ring.
The endomorphism $1$ corresponds to the identity in the Jacobi ring and its trace
is zero, according to Eq.~(\ref{l9}). The endomorphism $\xi_1\xi_2$ (which
geometrically represents a basis element for $H^1(T)$) corresponds to the
element $x_1x_2x_3$ in the Jacobi ring, and its trace is a nonzero constant,
according to Eq.~(\ref{l9}). Thus we have recovered the correct Yukawa
couplings for the elliptic curve by regarding the closed-string algebra
as the \Ho\ cohomology of the category of D-branes for the corresponding LG orbifold.

\subsection{Bulk-boundary maps}

Let us consider the simplest open-closed correlator: 2-point function on a disk,
with one boundary and one bulk insertion. It can be regarded as a linear function
$$
\xAop\ot \xAcl\ra \CC.
$$
Using the topological metric on $V_o$, it can also be regarded as a map
$$
\iota: \xAcl\ra \xAop.
$$
This is known as the bulk-boundary map; it is the dual of the boundary trace $\trb$.
Of course, for every choice of a D-brane,
we get a $\xAop$, and therefore there are as many maps $\iota$ as there are D-branes.
Axioms of Topological Field Theory with boundaries require $\iota$ to
be an algebra homomorphism. In this subsection we will demonstrate that the bulk-boundary
map comes for free if we identify $\xAcl$ with the \Ho\ cohomology of the category of
topological D-branes. We will also check in some examples that the resulting bulk-boundary
map is the correct one, i.e. it coincides with the one derived by physical methods.

Let $\cA=(A,Q,B)$ be a CDG algebra, and suppose the category of D-branes is the derived category
of CDG modules over $\cA$. (The case of DG algebras and modules is a special case of this,
corresponding to $B=0$.) Consider\footnote{The following construction was explained to us
by D.~Orlov.} the \Ho\ cohomology of $\cA$, defined as
$$
\HH^*(\cA)=\Ext^*_{\cA\ot\cA^{op}}(\cA,\cA).
$$
That is, it is the endomorphism algebra of $\cA$ regarded as an object of the derived category
of modules over $\cA\ot\cA^{op}$. Let $\cM$ be some object of the derived category of modules
over $\cA$, and let $\phi$ be an endomorphism of $\cM$. Then for any element $\alpha\in \HH^*(\cA)$
we may consider $\alpha\ot\phi$, which is an endomorphism of
$$
\cA\ot_{\cA} \cM\in D^b(\cA).
$$
But the latter object is isomorphic to $\cM$. Hence we may regard $\alpha\ot\phi$ as
as endomorphism of $\cM$. Thus we get a linear map
$$
\kappa_\cM: \HH^*(\cA)\ot \End(\cM)\ra \End(\cM).
$$
Obviously, this makes $\End(\cM)$ a (left) graded module over $\HH^*(\cA)$. Similarly, one can define
the right $\HH^*(\cA)$-module structure on $\End(\cM)$. More generally, if we consider any two objects
$\cM$ and $\cN$, the space of morphisms between them has a natural structure of a graded bimodule
over $\HH^*(\cA)$. We define the bulk-boundary map $\iota_\cM$ to be
\begin{equation}\label{defiota}
\iota_\cM (\alpha)=\kappa_\cM (\alpha\ot 1).
\end{equation}
It is trivial to see that this is an algebra homomorphism, as required.

Let us perform a simple check of this prescription for defining $\iota$. In the case
of a LG model on an affine space $V$, the disk correlator has been computed in Ref.~\cite{KLi2},
and the corresponding bulk-boundary map is given by
\begin{equation}\label{LGiota}
\iota_\cM: f\mapsto f\ \id_M
\end{equation}
where $f$ is an element of the Jacobi ring $\JRW=\CC[V]/\partial W$, and $\id_M$ is the identity endomorphism of a free CDG-module $\cM=(M,D)$ over the CDG algebra $(\CC[V],0,W)$. It is implicit
in this formula that $f$ is lifted to a polynomial on $V$. Although lifting an element of
the Jacobi ring to an element of $\CC[V]$ involves a choice, the above formula for $\iota_\cM$
is well-defined. Indeed, if we replace $f$ by
$$
f+\sum_{i=1}^n g_i\partial_i W
$$
where $g_i$ are some polynomials on $V$, then $\iota_\cM(f)$ will change by
$$
\left[D,\ \sum_i g_i \partial_i D\right],
$$
which is homotopic to zero (or in other words, is BRST-trivial).

On the other hand, we established in Section 3 that the Jacobi ring is isomorphic
to the endomorphism algebra of the ``diagonal brane'' on $V\times V$, the isomorphism
being given by
$$
f\mapsto f\ \id_R,
$$
where $R$ is the twisted Koszul resolution of the diagonal in $V\times V$. From this
it follows that the $\HH^*(\cA)$-module structure on $\End(\cM)$ is very simple:
it is induced by the $A$-module structure on $\End_A(M)$. Hence the abstract map
$\iota_\cM$ defined in Eq.~(\ref{defiota}) coincides with Eq.~(\ref{LGiota}).

\section{Discussion}

In this paper we have shown that for topological Landau-Ginzburg models on the affine space
(as well as on some quotients of the affine space) the algebra
of physical closed-string states\footnote{We are talking about gravitational primaries
only.} is isomorphic to the \Ho\ cohomology of the category
of D-branes. The algebra structure on the \Ho\ cohomology is given by the
standard cup product. Moreover, using the open-string topological metric (in mathematical
terms, the Serre functor on the category of D-branes), one can define a scalar product
on the \Ho\ cohomology, making it into a supercommutative Frobenius algebra. We showed
that this scalar product agrees with the closed-string topological metric.

An important question is whether more complicated correlators can be deduced by studying
the \Ho\ cohomology of the category of D-branes. This is not completely trivial even at tree level
(i.e. in genus zero). While two- and three-point correlators of gravitational
primaries in genus zero are encoded in the Frobenius algebra structure,
higher-point correlators can be described by a germ of a Frobenius manifold (whose
tangent space at the base point is the Frobenius algebra just mentioned.)
In the case of the B-model of a Calabi-Yau manifold, the corresponding
formal Frobenius manifold was constructed in Ref.~\cite{BK}. Frobenius manifolds
related to Landau-Ginzburg models are discussed in many papers, see e.g. Ref.~\cite{Man3c}.
It would be interesting to understand how to construct all these Frobenius
manifolds in a uniform manner, starting from natural algebraic structures on the \Ho\ complex of a category of topological D-branes. The key property needed
for such a construction seems to be the existence of the open-string metric,
i.e. a trivial Serre functor, on the category of D-branes.

A further question is how to reconstruct open-closed correlators. We only
considered the simplest example (disk correlator with one boundary and one bulk insertion)
and showed how to recover it from the $\HH^*(\cA)$-module structure on the space of endomorphisms
of any D-brane.

Although in this paper we have focused on topological string theories, it is tempting
to conjecture that something similar holds for bosonic strings.
A matter CFT with central charge $c=26$ coupled to diffeomorphism ghosts can be
thought of as a topological field theory~\cite{topgravDist,Li,GRSe,Bersh,MV,WiCS}
and can be coupled to topological gravity in the usual manner. This gives an alternative description of the
bosonic string, which is similar to that of topological string theory. However, the physical spectrum of the bosonic
string theory is related not to the BRST cohomology of the TFT, but to
its semi-relative cohomology~\cite{Zwiebach,WiCS,Getz2}. In particular,
one cannot classify physical states as being gravitational primaries or descendants,
in general, and it is not clear what the relation between physical closed string
states and the \Ho\ cohomology could be.
A good example to look at is noncritical bosonic string theory with $c<1$.
Some of the bosonic $c<1$ backgrounds are equivalent in a nontrivial way to
topological string theories of the kind discussed in this paper~\cite{topgravWi,topgravDist,DiWi,Li},
and it would be interesting to rephrase our results in terms of the conventional
formulation of the bosonic string.

\section*{Acknowledgments}
A.K. would like to thank Volodya Baranovsky, Ezra Getzler, Kentaro Hori, Dima Orlov, and
Sasha Voronov for help at various stages. A.K. is also grateful to the
Department of Mathematics of Northwestern University and the Erwin Schr\"odinger
Institute for hospitality while this work was being completed.
L. R. is very grateful to Mikhail Khovanov for numerous discussions of
the category of matrix factorizations.
This work was supported in part by the DOE grant
DE-FG03-92-ER40701 and by the NSF grant DMS-0196131.


\begin{thebibliography}{99}

\bibitem{Diac}
S.~K.~Ashok, E.~Dell'Aquila and D.~E.~Diaconescu, {\em Fractional branes in Landau-Ginzburg orbifolds,}
arXiv:hep-th/0401135.

\bibitem{BK}
S.~Barannikov and M.~Kontsevich, {\em Frobenius manifolds and formality of Lie
algebra of polyvector fields,} Internat.~Math.~Res.~Notices no. 4 (1998) 201 [arXiv:alg-geom/9710032].


\bibitem{Bersh}
M.~Bershadsky, W.~Lerche, D.~Nemeschansky and N.~P.~Warner,
{\em Extended N=2 superconformal structure of gravity and W gravity coupled to
matter,} Nucl.\ Phys.\ B {\bf 401}, 304 (1993)
[arXiv:hep-th/9211040].

\bibitem{Bru1}
I.~Brunner, M.~R.~Douglas, A.~E.~Lawrence and C.~Romelsberger, {\em D-branes on the quintic,}
JHEP {\bf 0008}, 015 (2000)
[arXiv:hep-th/9906200].

\bibitem{Letal}
I.~Brunner, M.~Herbst, W.~Lerche and B.~Scheuner,
{\it Landau-Ginzburg realization of open string TFT,}
arXiv:hep-th/0305133.

\bibitem{Bru2}
I.~Brunner and V.~Schomerus, {\em On superpotentials for D-branes in Gepner models,}
JHEP {\bf 0010}, 016 (2000)
[arXiv:hep-th/0008194].

\bibitem{DD}
D.~E.~Diaconescu and M.~R.~Douglas, {\em D-branes on stringy Calabi-Yau manifolds,}
arXiv:hep-th/0006224.


\bibitem{DiWi}
R.~Dijkgraaf and E.~Witten,
{\em Mean Field Theory, Topological Field Theory, And Multimatrix Models,}
Nucl.\ Phys.\ B {\bf 342}, 486 (1990).

\bibitem{topgravDist}
J.~Distler, {\em 2-D Quantum Gravity, Topological Field Theory And The Multicritical Matrix
Models,} Nucl.\ Phys.\ B {\bf 342}, 523 (1990).

\bibitem{GRSe}
B.~Gato-Rivera and A.~M.~Semikhatov,
{\em Minimal models from W constrained hierarchies via the Kontsevich-Miwa
transform,} Phys.\ Lett.\ B {\bf 288}, 38 (1992)
[arXiv:hep-th/9204085].


\bibitem{Gepner}
D.~Gepner, {\em Exactly Solvable String Compactifications On Manifolds Of SU(N) Holonomy,}
Phys.\ Lett.\ B {\bf 199}, 380 (1987).

\bibitem{Ge}
M.~Gerstenhaber, {\em The cohomology structure of an associative ring,}
Ann. of Math. {\bf 2} (1963) 267.

\bibitem{GS}
M.~Gerstenhaber and S.~D.~Schack, {\em Algebraic cohomology and deformation theory,}
in: Deformation theory of algebras and structures and applications (Il Ciocco, 1986),
pp. 11-24, Kluwer, 1988.

\bibitem{Getz1}
E.~Getzler,
{\em Batalin-Vilkovisky Algebras And Two-Dimensional Topological Field Theories,}
Commun.\ Math.\ Phys.\  {\bf 159}, 265 (1994)
[arXiv:hep-th/9212043].

\bibitem{Getz2}
E.~Getzler,
{\em Two-Dimensional Topological Gravity And Equivariant Cohomology,}
Commun.\ Math.\ Phys.\  {\bf 163}, 473 (1994)
[arXiv:hep-th/9305013].



\bibitem{GutSa}
M.~Gutperle and Y.~Satoh, {\em D-branes in Gepner models and supersymmetry,}
Nucl.\ Phys.\ B {\bf 543}, 73 (1999)
[arXiv:hep-th/9808080].

\bibitem{HerbstLaz}
M.~Herbst and C.~I.~Lazaroiu, ``Localization and traces in open-closed topological Landau-Ginzburg models,''
arXiv:hep-th/0404184.

\bibitem{Lazrec}
M.~Herbst, C.~I.~Lazaroiu and W.~Lerche,
{\em Superpotentials, $A_\infty$ relations and WDVV equations for open
topological strings,} arXiv:hep-th/0402110.

\bibitem{Horec}
K.~Hori and J.~Walcher,
{\em F-term equations near Gepner points,}
arXiv:hep-th/0404196.

\bibitem{IV}
K.~A.~Intriligator and C.~Vafa, {\em Landau-Ginzburg Orbifolds,}
Nucl.\ Phys.\ B {\bf 339}, 95 (1990).

\bibitem{KLi1}
A.~Kapustin and Y.~Li,
{\em D-branes in Landau-Ginzburg models and algebraic geometry,}
JHEP {\bf 0312}, 005 (2003)
[arXiv:hep-th/0210296].


\bibitem{KLi2}
A.~Kapustin and Y.~Li,
{\em Topological correlators in Landau-Ginzburg models with boundaries,}
arXiv:hep-th/0305136.

\bibitem{KatzSharpe} S.~Katz and E.~Sharpe, {\em D-branes, open string vertex operators, and Ext groups,} arXiv:hep-th/0208104.

\bibitem{KhRo}
M.~Khovanov and L.~Rozansky, {\em Matrix factorizations and link homology,}
arXiv:math.QA/0401268

\bibitem{Konts}
M.~Kontsevich, {\em Homological Algebra of Mirror Symmetry}, arXiv:alg-geom/9411018.

\bibitem{KKS}
T.~Kugo, H.~Kunitomo and K.~Suehiro, {\em Nonpolynomial Closed String Field Theory,}
Phys.\ Lett.\ B {\bf 226}, 48 (1989).

\bibitem{Lazaxi} C.~I.~Lazaroiu, {\em On the structure of open-closed topological field theory in two  dimensions}, Nucl.\ Phys.\ B {\bf 603}, 497 (2001) [arXiv:hep-th/0010269].

\bibitem{Laz}
C.~I.~Lazaroiu, {\it On the boundary coupling of topological Landau-Ginzburg models,}
arXiv:hep-th/0312286.

\bibitem{Li}
K.~Li, {\em Topological Gravity With Minimal Matter,}
Nucl.\ Phys.\ B {\bf 354}, 711 (1991).

\bibitem{LZ}
B.~H.~Lian and G.~J.~Zuckerman,
{\em New perspectives on the BRST algebraic structure of string theory,}
Commun.\ Math.\ Phys.\  {\bf 154}, 613 (1993)
[arXiv:hep-th/9211072].

\bibitem{LZ2}
B.~H.~Lian and G.~J.~Zuckerman,
{\em Some Classical And Quantum Algebras,}
arXiv:hep-th/9404010.

\bibitem{Man3c}
Yu.~I.~Manin, {\em Three constructions of Frobenius manifolds: a comparative study,}
in: Sir Michael Atiyah, a great mathematician of the twentieth century, Asian J. of
Math. {\bf 3} (1999) 179 [arXiv:math.QA/9801006].


\bibitem{MV}
S.~Mukhi and C.~Vafa,
{\em Two-dimensional black hole as a topological coset model of c = 1 string
theory,} Nucl.\ Phys.\ B {\bf 407}, 667 (1993)
[arXiv:hep-th/9301083].


\bibitem{PS}
M.~Penkava and A.~Schwarz, {\em $A_\infty$ Algebras And The Cohomology Of Moduli Spaces,}
arXiv:hep-th/9408064.

\bibitem{PS2}
M.~Penkava and A.~Schwarz,
{\em On some algebraic structure arising in string theory,}
arXiv:hep-th/9212072.


\bibitem{RS}
A.~Recknagel and V.~Schomerus, {\em D-branes in Gepner models,}
Nucl.\ Phys.\ B {\bf 531}, 185 (1998)
[arXiv:hep-th/9712186].

\bibitem{SaaZwi}
M.~Saadi and B.~Zwiebach, {\em Closed String Field Theory From Polyhedra,}
Annals Phys.\  {\bf 192}, 213 (1989).

\bibitem{Swan}
R.~G.~Swan, {\it \Ho\ cohomology of quasiprojective schemes,} J. Pure Appl. Algebra
{\bf 110} (1996) 57.

\bibitem{Vafa}
C.~Vafa, {\em Topological Landau-Ginzburg Models,}
Mod.\ Phys.\ Lett.\ A {\bf 6}, 337 (1991).



\bibitem{WeiGel}
C.~A.~Weibel and S.~C.~Geller, {\em \'{E}tale descent for \Ho\ and cyclic homology,}
Comment. Math. Helv. {\bf 66} (1991) 368.


\bibitem{Witten}
E.~Witten, {\em Noncommutative Geometry And String Field Theory,}
Nucl.\ Phys.\ B {\bf 268}, 253 (1986).

\bibitem{topgravWi}
E.~Witten,
{\em On The Structure Of The Topological Phase Of Two-Dimensional Gravity,}
Nucl.\ Phys.\ B {\bf 340}, 281 (1990).

\bibitem{WiCS}
E.~Witten, {\em Chern-Simons gauge theory as a string theory,}
Prog.\ Math.\  {\bf 133}, 637 (1995)
[arXiv:hep-th/9207094].


\bibitem{phases}
E.~Witten, {\em Phases of N = 2 theories in two dimensions,}
Nucl.\ Phys.\ B {\bf 403}, 159 (1993)
[arXiv:hep-th/9301042].

\bibitem{WZ}
E.~Witten and B.~Zwiebach, {\em Algebraic structures and differential geometry in $2D$ string theory,}
Nucl.\ Phys.\ B {\bf 377}, 55 (1992) [arXiv:hep-th/9201056].

\bibitem{Zwiebach}
B.~Zwiebach, {\em Closed string field theory: Quantum action and the B-V master equation,}
Nucl.\ Phys.\ B {\bf 390}, 33 (1993)
[arXiv:hep-th/9206084].


\end{thebibliography}
\end{document}